\documentstyle[epsfig,twocolumn]{elsart}
\begin{document}
\begin{frontmatter}

\title{Inhomogeneous superconductivity induced in a weak ferromagnet}
\author[Leiden]{A.Rusanov\thanksref{cor-aut}},
\author[Leiden]{R.Boogaard},
\author[Leiden]{M.Hesselberth},
\author[Grenoble]{H. Sellier}
\author[Leiden]{J.Aarts}
\thanks[cor-aut]{Corresponding author; email : rusanov@phys.leidenuniv.nl}

\address[Leiden]{Kamerlingh Onnes Laboratory, Leiden University, P.O. Box 9506,
 2300 RA Leiden, The Netherlands}
\address[Grenoble]{Departement de Recherche Fondamentale sur la Matiere
Condensee SPSMS/LCP, CEA/Grenoble, France}

\begin{abstract}
Under certain conditions, the order parameter induced by a
superconductor (S) in a ferromagnet (F) can be inhomogeneous and
oscillating, which results e.g. in the so-called $\pi$-coupling in
S/F/S junctions. In principle, the inhomogeneous state can be
induced at $T_c$ as function of the F-layer thickness $d_F$ in S/F
bilayers and multilayers, which should result in a dip-like
characteristic of $T_c(d_F)$. We show the results of measurements
on the S/F system Nb/Cu$_{1-x}$Ni$_x$, for Ni-concentrations in
the range $x$~= 0.5-0.7, where such effects might be expected. We
find that the critical thickness for the occurrence of
superconductivity is still relatively high, even for these weak
ferromagnets. The resulting dip then is intrinsically shallow and
difficult to observe, which explains the lack of a clear signature
in the $T_c(d_F)$ data. \\ $\newline$ Keywords: Proximity effect,
SFS-junctions, LOFF-state
\end{abstract}
\end{frontmatter}

\small

\section{Introduction}
Recently it was shown that
superconductor/ferromagnet/superconductor (S/F/S) junctions made
of $Nb/Cu_{1-x}Ni_x/Nb$, where the interlayer is weakly
ferromagnetic ($x \approx$ 0.54), can support a supercurrent.
Moreover, the temperature dependence of the supercurrent shows a
sharp cusp, which suggests that the junction changes from a
0-phase to a $\pi$-phase state at low temperatures \cite{ryaz01}.
This implies that the superconducting order parameter induced in
the ferromagnet is oscillatory damped, and also that the
transparency of the interface is relatively high. A signature for
this inhomogeneous state should also be visible in the
T$_c$-dependence of S/F multilayers as the function of the F-layer
thickness d$_F$ \cite{rado91,tagir98}. Specifically, T$_c$ should
go through a dip before reaching a maximum and going to an
asymptotic value, under the condition, however, that the S-layer
thickness d$_S$ is of the order of the superconducting coherence
length $\xi_S$. The dip, which appears to be a stronger signature
for the inhomogeneous state than the often searched-for maximum
(see e.g \cite{laz00,aarts97}), signifies a self-interference
effect of the inhomogeneous order parameter, and could be used to
advantage in constructing a superconducting spin switch
\cite{tagir99,buz99}. The necessary conditions can be reached if
the pair breaking by the ferromagnet is not too strong, so that
the critical thickness $d^S_{cr}$ of the superconductor is not
(much) larger than $\xi_S$. In principle, given the observation of
supercurrents in Nb/Cu$_{0.46}$Ni$_{0.54}$/Nb, this system might
also be a viable candidate for observing the interference effects.
Here we present results of the measurements of the magnetic and
superconducting properties of Nb/Cu$_{1-x}$Ni$_x$ for different
$x$. We show that $d^S_{cr}$ even for these weak ferromagnets is
high enough to make the predicted effects small.

\section{Sample preparation and magnetic properties}
\begin{figure}
\epsfig{file=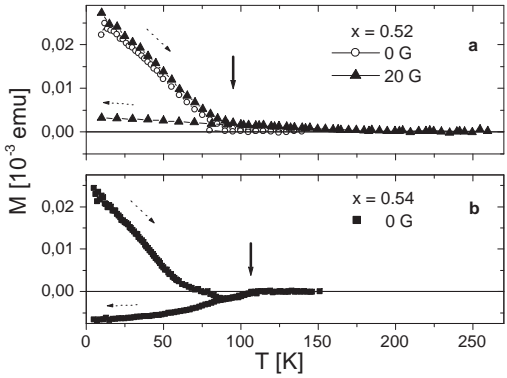,width=\columnwidth}
\caption{Magnetization $M$ versus temperature $T$ for single films
of Cu$_{1-x}$Ni$_x$ with (a) $x$ = 0.52 (applied field 0 G, 20 G),
(b) $x$ = 0.54 (0 G). Dotted arrows show the measurement sequence,
solid arrows the values of $T_{Curie}$.} \label{fig:fig1}
\end{figure}
Sets of single F-films, S/F bilayers and F/S/F trilayers were
DC-magnetron sputtered in an ultra high vacuum system with base
pressure 10$^{-9}$ mbar and sputtering argon pressure about
6*10$^{-3}$ mbar. Cu$_{1-x}$Ni$_x$ targets were used with $x$~=
0.60, 0.50 and 0.45 atomic percent which yielded a somewhat
different Ni concentration in the samples~: $x$~= 0.67, 0.59 and
0.52 respectively. The samples were measured by SQUID-magnetometry
in order to determine the saturation magnetization $M_s$ and the
Curie temperature $T_{Curie}$. For comparison we also measured a
set of samples which were sputtered in an RF-sputtering system
with a base pressure of 10$^{-7}$ mbar \cite{sput}.
\begin{figure}
\epsfig{file=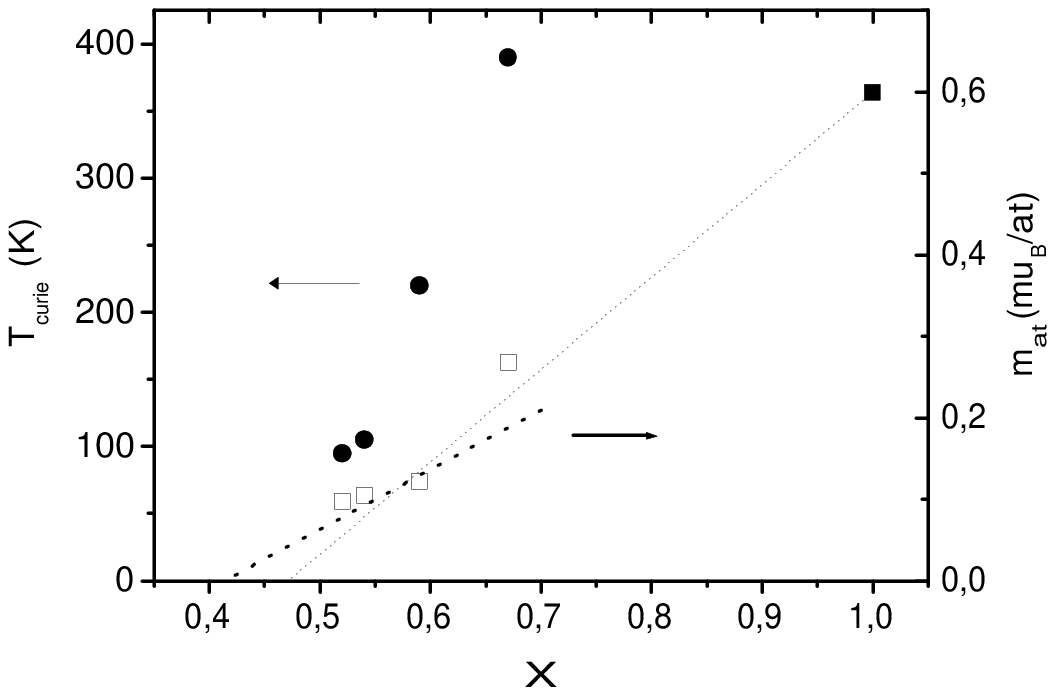,width=\columnwidth}
\caption{Ferromagnetic transition temperature  $T_{Curie}$ and
magnetic moment per atom (formula unit) $m_{at}$ as function of
concentration $x$ for alloy films of Cu$_{1-x}$Ni$_x$. The dotted
lines show the behavior of the bulk magnetic moment
\cite{ahern58}. The kink in the functional dependence is
emphasized by extrapolating the two linear regimes. The solid
square shows the bulk moment for pure Ni.}
 \label{fig:fig2}
\end{figure}
 Fig.~\ref{fig:fig1} shows the typical dependence of the
magnetization $M$ on temperature $T$ for a single F-layer, using
the following procedure: the sample was magnetized to its
saturation at 10 K, then the field was removed (or set to a small
value), and $M(T)$ was measured up to 300 K and back down to 10 K.
$T_{Curie}$ was defined at the temperature where $M(T)$ deviates
from the constant value at high $T$, which is usually slightly
higher than where the hysteresis sets in. The negative part of the
$M(T)$ curve in Fig.~\ref{fig:fig1}b we associate with a small
negative residual field in the cryostat. Fig.~\ref{fig:fig2} and
Table~1 show T$_{Curie}$ and the magnetic moment $\mu_{at}$ of the
CuNi films as function of Ni concentration. In the range $x$~=
0.52 - 0.59, $\mu_{at}$ changes very little, with a stronger
increase above $x$~= 0.6. Also shown in Fig.~\ref{fig:fig2} is the
behavior of the bulk magnetic moments according to
ref.~\cite{ahern58}. The agreement is quite satisfactory and the
comparison makes clear that the small changes below $x$~= 0.6
accurately mimic the bulk behavior, where a kink in the linear
dependence is found at that value. Interestingly, T$_{Curie}(x)$
behaves somewhat differently, with a much larger variation. It
suggests that T$_{Curie}$ is a more sensitive measure for $x$ than
$\mu_{at}$.

\section{Superconducting properties and discussion}

\begin{figure}
\epsfig{file=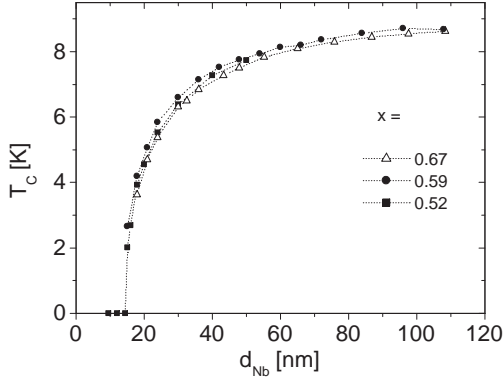,width=\columnwidth} \caption{Critical
temperature $T_c$ versus Nb thickness $d_{Nb}$ for F/S/F trilayers
with F = Cu$_{1-x}$Ni$_x$, S = Nb, for $x$~= 0.67, 0.59 and 0.52.
Dotted lines are meant to guide the eye.} \label{fig:fig3}
\end{figure}
The dependence of the critical temperatures T$_c$ on d$_{Nb}$ of
Cu$_{1-x}$Ni$_x$/Nb/Cu$_{1-x}$Ni$_x$ sandwiches with thick
F-layers (50 nm)  is shown in Fig.~\ref{fig:fig3} for three Ni
concentration: $x$~= 0.52, 0.59 and 0.67. Values for $d^S_{cr}$
for all concentrations are presented in Table~1. There is hardly
any dependence on Ni concentration. The data for $x$~= 0.67 yield
a slightly higher value for d$^S_{cr}$ while the data for $x$~=
0.59 even lie on the low-$d_{Nb}$ side of the data for $x$~= 0.52,
but for all sets measured, d$^S_{cr}$ is about 14 nm. In view of
the small changes in $\mu_{at}$ found above, this indicates that
$\mu_{at}$ is a better measure for the pair breaking effects in
the F-layer than $T_c$. Using the value $\xi_S$~= 8 nm for Nb
\cite{aarts97} we find $d^S_{cr}$/$\xi_S\approx$ 1.6. This already
implies that the signature of the inhomogeneous superconductivity,
namely the dip in T$_c$(d$_F$), can only be small \cite{tagir98}.

For T$_c$(d$_F$), measurements were performed on two sets of S/F
bilayers (S-layer on the substrate side), one with (S) $d_{Nb}$ =
18 nm and (F) $x$~= 0.59, the other with (S) $d_{Nb}$~= 12 nm and
(F) $x$~= 0.52.
\begin{figure}
\epsfig{file=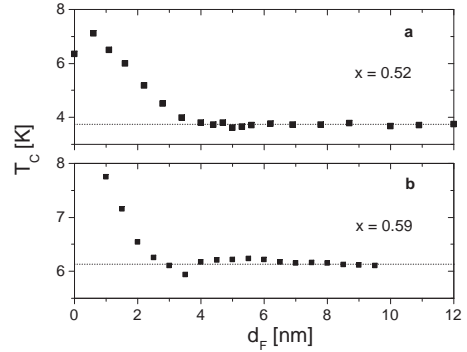,width=\columnwidth} \caption{Critical
temperature $T_c$ versus F-layer thickness $d_F$ for S/F bilayers
with F =  Cu$_{1-x}$Ni$_x$, S = Nb. (a) $x$~= 0.59, $d_{Nb}$~= 18
nm; (b) $x$~= 0.52, $d_{Nb}$~= 12 nm.The dotted lines indicate the
asymptotic value at large $d_F$. } \label{fig:fig4}
\end{figure}
The data in Fig.~\ref{fig:fig4}a,b show a possible dip in the case
of $x$~= 0.59, but only a kink in the case of $x$~= 0.52. Given
the small difference in $\mu_{at}$ the similar behavior is not
surprising. The absence of (stronger) dips is probably due to a
combination of the still rather large value for $d^S_{cr}$/$\xi_S$
and bandstructure effects, which make the interfaces less than
fully transparent. Also, growth conditions are very important, and
smearing effects likely. A final remark concerns the value of
$d_F$ where the kink is found, around 4~- 5~nm. It should be
realized that this does not contradict the coupling thickness for
reaching the $\pi$-state in S/F/S junctions of 15-20 nm : firstly,
the $T_c(d_F)$ data are on bilayers and therefore correspond to
8~- 10~nm for the trilayer case; secondly because the weak
magnetism results in a temperature dependence of $\xi_F$
\cite{ryaz01}.
\begin{center}
\begin{table}[t]
\begin{tabular}{|c|ccc|} \hline
 $x$ &  $d^S_{cr} / \xi_S$& $T_{Curie}$ [K]&
 $\mu_{at}$ [$\mu_B$/at] \\ \hline
 0.67 & 1.63 & 390 & 0.27 \\
 0.59 & 1.55 & 220 & 0.12 \\
 0.54 &      & 105 & 0.11 \\
 0.52 & 1.55  &  95 & 0.10 \\ \hline
\end{tabular}
\vspace{2mm} \caption{Values of $d^S_{cr}$ in units of $\xi_S$ for
F/Nb/F trilayers (F = Cu$_{1-x}$Ni$_x$) and of $T_{Curie}$ and the
magnetic moment per atom $\mu_{at}$ as determined on single F
films for different Ni-concentrations of the ferromagnetic
Cu$_{1-x}$Ni$_x$, as indicated.}
\end{table}
\end{center}

\section*{Acknowledgements}
This work is part of the research program of the 'Stichting voor
Fundamenteel Onderzoek der Materie (FOM)', which is financially
supported by NWO. We acknowledge stimulating discussions with L.
Tagirov and A. Golubov. H. S. acknowledges a visiting grant from
the European cooperation in the field of Scientific and Technical
research (COST), action P5 (mesoscopic electronics).

\end{document}